\documentclass{PoS}

\usepackage[scaled=0.9]{helvet}

\usepackage{cite}
\usepackage{url}
\usepackage{amsmath,array}

\def\ibar{\overline I_0}
\def\gev{\,\mathrm{GeV}}
\def\mev{\,\mathrm{MeV}}
\def\n#1e#2n{#1\times 10^{#2}}
\def\mbstar{m_{B^*}}

\def\sth{s_\mathrm{th}}
\def\qsqmax{q^2_\mathrm{max}}
\def\kl#1{K_{l#1}}
\def\modvub{|V_{ub}|}

\title{Elastic $s$-wave scattering phase shifts and $|V_{ub}|$ from
lattice calculations of form factors for exclusive semileptonic
decays}

\ShortTitle{Phase shifts and $|V_{ub}|$ from exclusive semileptonic
decays}

\author{\speaker{J.M. Flynn}\\%
        University of Southampton, UK\\
        E-mail: \email{jflynn@phys.soton.ac.uk}}

\author{J. Nieves\\
        Universidad de Granada, Spain\\
        E-mail: \email{jmnieves@ugr.es}}

\abstract{Omn\`es dispersion relations make a connection between form
  factors for exclusive semileptonic decays and phase shifts in the
  corresponding elastic scattering channels. We describe two
  applications. In the first, we use lattice form factor calculations
  to learn about phase shifts in elastic $s$-wave isospin-$1/2$
  channels for $K\pi$, $B\pi$, $D\pi$ and $DK$ scattering. The aim of
  the second application is to make the determination of the CKM
  matrix element magnitude $|V_{ub}|$ from exclusive semileptonic
  $B\to\pi$ decays competitive with that from inclusive decays. Here
  we use many subtractions in an Omn\`es dispersion relation to
  motivate a simple fitting function, allowing data to constrain the
  $q^2$ shape of the differential decay rate and theory, primarily
  lattice results, to provide normalisation via form factor values.}

\FullConference{The XXV International Symposium on Lattice Field Theory\\
		 July 30 -- August 4 2007\\
		 Regensburg, Germany}

\begin{document}

\section{Omn\`es dispersion relations}

Mandelstam's hypothesis of maximum analyticity and Watson's Theorem
relate the phases of the form factors $f$ in exclusive semileptonic
$M\to\pi$ decay (where $M \in\{B,D,K\}$) to the phase shifts in the
elastic $M\pi\to M\pi$ scattering amplitudes in the corresponding
$J^P$ and isospin channels. We have
\begin{equation}
\frac{f^+(s+\mathrm{i}\epsilon )}{f^+(s-\mathrm{i}\epsilon)} =
\frac{T(s+\mathrm{i}\epsilon )}{T(s-\mathrm{i}\epsilon )} =
e^{\,2i \delta(s) },\qquad s > \sth
\end{equation}
where $\sth=(m_M+m_\pi)^2$ and $T(s)$ is the scattering amplitude,
related to the phase shift $\delta(s)$ by
\begin{equation}
\label{eq:Tdelta}
T(s) = \frac{8\pi\mathrm{i}s}{\lambda^{1/2}(s)}
 (e^{\,2\mathrm{i}\delta(s)}-1)
\end{equation}
where $\lambda$ is the usual kinematic function. The (inverse)
scattering amplitude, in the appropriate isospin and angular momentum
channel, is found from~\cite{EJ99,EJ99_2}
\begin{equation}
\label{eq:Tinverse}
T^{-1}(s) = -\ibar(s)-\frac1{8\pi a\sqrt\sth}
          +\frac1{V(s)}-\frac1{V(\sth)}
\end{equation}
Here, $V$ is the two-particle irreducible scattering amplitude, $a$ is
the scattering length and $\ibar$ is calculated from a one-loop bubble
diagram. This description automatically implements elastic unitarity,
which is necessary for the phase shift to be extracted from
equation~(\ref{eq:Tdelta}).

For multiple multiple subtractions, $\{(q_i^2,f_i): i=0,\dots,n\}$,
the Omn\`es result reads
\begin{align}
f(q^2) &=
  \prod_{j=0}^n f_j^{\alpha_j(q^2)} \times
  \exp\Big\{I_{\!\delta}(q^2;\{q_j^2\})
  \prod_{k=0}^n(q^2{-}q^2_k)\Big\} \\
I_{\!\delta}(q^2;\{q^2_j\}) &=
  \frac{1}{\pi}\int_{\sth}^{+\infty}\!\!\!
  \frac{ds}{(s{-}q^2_0)\cdots(s{-}q^2_n)}\frac{\delta(s)}{s-q^2}
\end{align}
\begin{equation}
\alpha_j(q^2) =
  \prod_{k=0,\,\, k\neq j}^{n} \frac{q^2-q^2_k}{q^2_j-q^2_k}, \qquad
\alpha_j(q_i^2) = \delta_{ij}, \qquad
\sum_{k=0}^n \alpha_k(q^2)  =  1
\end{equation}
One can balance the number of subtractions against knowledge of
$\delta$. In the first application below we use (one or) two
subtractions and form-factor input information to extract the
scattering length in the corresponding elastic scattering
channels~\cite{Flynn:2007ki}. In the second application we make many
subtractions to motivate a simple parametrisation of the form factors
for exclusive semileptonic $B\to\pi$ decays, allowing the extraction
of $\modvub$ from lattice form factor results combined with
experimental partial branching fraction
information~\cite{Flynn:2007ii,Flynn:2007qd}.

\section{Elastic $s$-wave $K\pi$, $B\pi$, $D\pi$ and $DK$ scattering
  lengths}

We use lattice calculations of the scalar form factor $f_0(q^2)$ in
exclusive semileptonic decays for input. In the Omn\`es dispersion
relation we use one or two subtractions to retain dependence on the
phase shift and apply lowest order chiral perturbation theory (ChPT)
or heavy meson chiral perturbation theory (HMChPT) for the
two-particle irreducible amplitudes $V$ needed for
equation~(\ref{eq:Tinverse}). In our fits we can then determine the
scattering length, $m_\pi a$, and the form factor values, $f_0$, at
the chosen subtraction points.

\subsection{Elastic $s$-wave $K\pi$ scattering}

\begin{figure}
\begin{center}
\includegraphics[width=0.49\hsize]{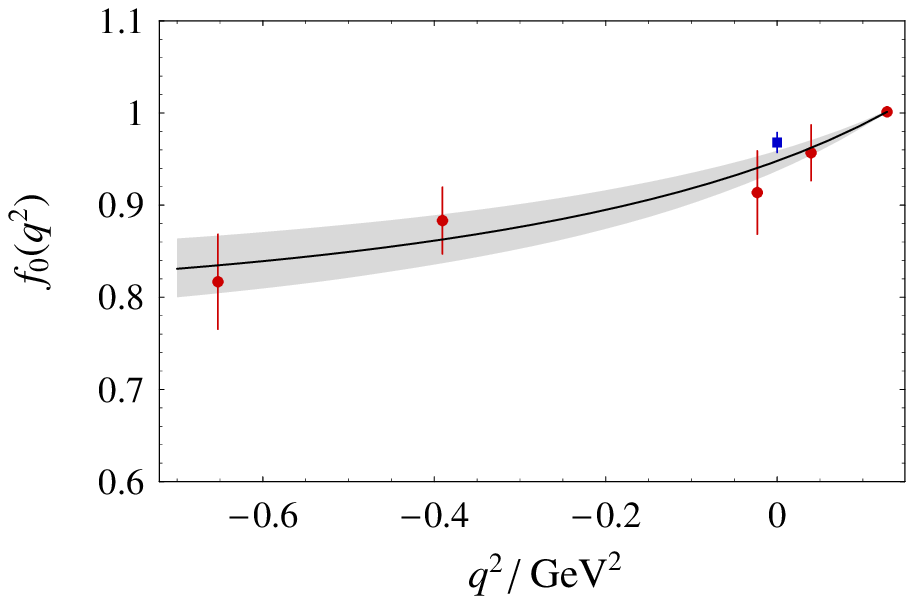}\hfill
\includegraphics[width=0.49\hsize]{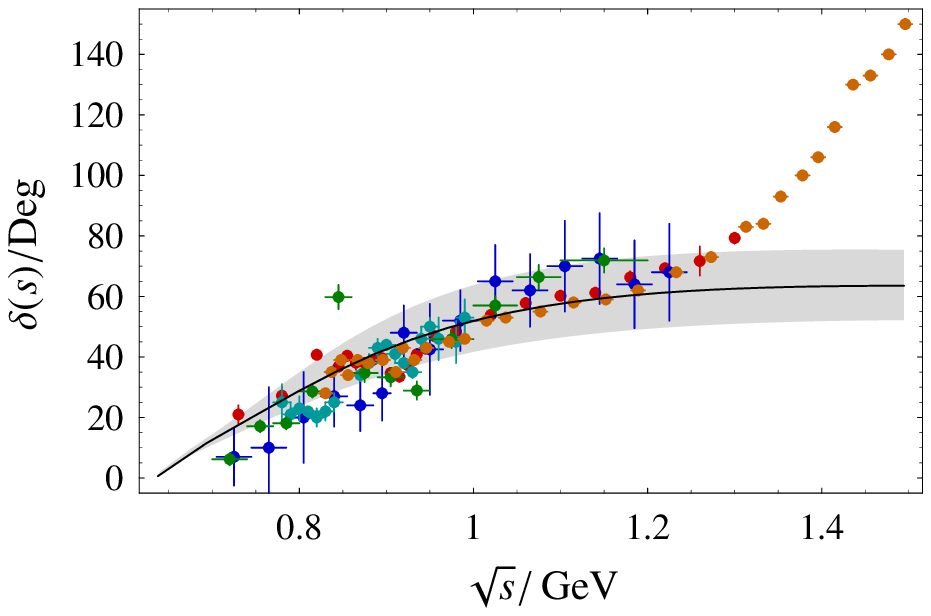}
\end{center}
\caption{The left hand plot shows the $\kl3$ form factor $f_0(q^2)$,
  with a $68\%$ error band, obtained from a fit using a
  twice-subtracted Omn\`es relation, implementing a linear relation
  between $f_0(0)$ and the scattering length as described in the text.
  Red points are the form-factor inputs and the blue square shows the
  result from~\cite{Dawson:2006qc} for $f_0(0)$ (not fitted). The
  right hand plot shows the isospin-$1/2$ $K\pi$ $s$-wave phase shift
  with a $68\%$ error band (grey). The phase shift plot also shows
  experimental data points
  from~\cite{Mercer:1971kn,Estabrooks:1977xe,Bingham:1972vy,Baker:1974kr,Aston:1987ir}.}
\label{fig:kl3-2sub-corr}
\end{figure}

For the isospin-$1/2$ scalar $K\pi$ channel, the lowest order ChPT
expression for $V$ is (with $f_\pi=92.4\mev$)
\begin{equation}
V(s) = \frac1{4f_\pi^2}
  \left( m_K^2+m_\pi^2 -\frac52 s + \frac3{2s}(m_K^2-m_\pi^2)^2
  \right).
\label{eq:scattamp}
\end{equation}
We take calculated values of the scalar form factor for $\kl3$ decays
from $N_f=2$ domain wall fermion results by RBC~\cite{Dawson:2006qc}.
Since this reference does not provide chirally-extrapopolated values
for the form factor except at $q^2=0$, we perform our own simple
chiral extrapolation, as described in~\cite{Flynn:2007ki}, to provide
input pairs $(q^2, f_0(q^2))$. To reduce the dependence on the phase
shift at large values of the centre-of-mass energy while retaining
sensitivity to the scattering length, we use subtraction points at
$q^2=0$ and $q_1^2=-0.75\gev^2$. Our two-subtraction fit shows almost
complete anticorrelation of $f_0(0)$ and the scattering length $m_\pi
a$, so we redo our fit, implementing a linear relation between them as
a constraint (we deduce the relation from a single-subtraction
fit)\footnote{The anticorrelation is not unexpected because the lowest
order ChPT expressions for $f_0(0)$ and the scattering length are
linearly related, depending only on $1/f_\pi^2$ (apart from masses).}.
Our results are:
\begin{equation}
f_0(q_1^2)=0.827(32), \quad
f_0(0) = 0.948(10), \quad
m_\pi a =  0.179(17)
\end{equation}
and our fitted form factor and phase shift are shown in
Figure~\ref{fig:kl3-2sub-corr}. The phase-shift plot also shows
experimental points for comparison: we emphasise that we have not fit
these data, so the agreement with the phase shift determined from a
lattice calculation is very encouraging.
Since the Omn\`es integration reaches values where massive resonance
exchanges could be relevant, we estimate the associated uncertainties
by incorporating the exchange of $\rho$ and $K^*$ resonances as well
as nonet scalar mesons with masses above $1\gev$, using the
isopsin-$1/2$ $K\pi$ scattering amplitude from~\cite{Jamin:2000wn}.
This also incorporates some next-to-leading ChPT effects. We find no
appreciable changes in the fitted form-factor values, while the
scattering length increases by $6\%$. We have also examined $K\eta$
coupled-channel effects finding again no appreciable changes in the
form-factor values and this time a decrease of up to $5\%$ in the
scattering length. Combining these effects, we ascribe an $8\%$
systematic error to the scattering length, leading to a result:
\begin{equation}
\label{eq:kpi-scattlength}
m_\pi a = 0.179(17)(14).
\end{equation}

\subsection{Elastic $s$-wave $B\pi$ scattering}

For the two-particle irreducible isospin-$1/2$ $s$-wave $B\pi$
scattering amplitude we use the leading contact term from the heavy
meson chiral perturbation theory (HMChPT)
lagrangian~\cite{Wise:1992hn},
\begin{equation}
\label{eq:hmchpt-amp}
V(s) = \frac1{4f_\pi^2}
 \left( 2(m_B^2+m_\pi^2)-3s +\frac{(m_B^2-m_\pi^2)^2}s
 \right).
\end{equation}
We have not included a contribution from the $t$-channel
$B^*$-exchange diagram depending on the leading HMChPT $B^*B\pi$
interaction term, since this vanishes at $\sth$ and has magnitude less
than $1\%$ of that from the expression above over a large range of
$s$.

We take input scalar form factor values from the lattice QCD
calculations by the HPQCD~\cite{Dalgic:2006dt} and
FNAL~\cite{Okamoto:2005zg} collaborations, assuming that the
statistical errors are uncorrelated, while the systematic errors are
fully-correlated. Note that the HPQCD results are updated from those
we used in~\cite{Flynn:2007ki}, while we have also added points read
off Figure~7 in~\cite{Okamoto:2005zg}. We also use the lightcone
sum rule result for $f_0(0)=f_+(0)$ from~\cite{LCSR_04_BZ}.

\begin{figure}
\begin{center}
\includegraphics[width=0.49\hsize]{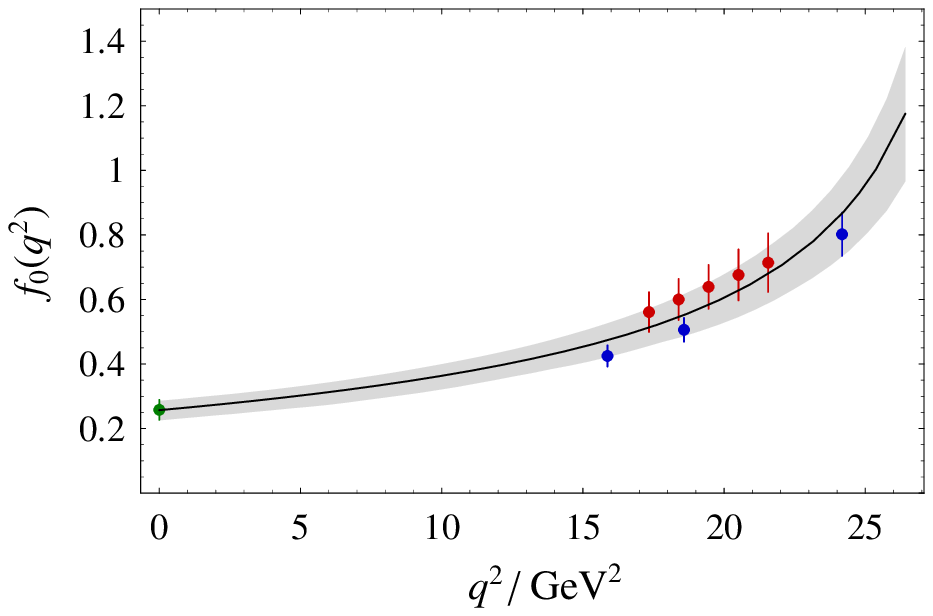}\hfill
\includegraphics[width=0.49\hsize]{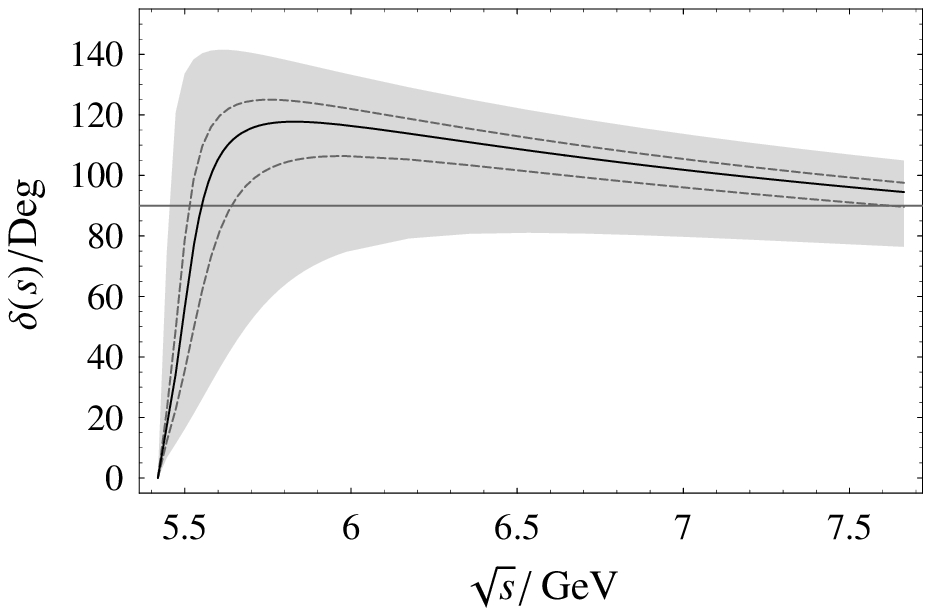}
\end{center}
\caption{$B\pi$ isospin-$1/2$ scalar form factor and phase shift,
together with $68\%$ confidence level bounds (grey bands). The points
on the form factor plot are the inputs
from~\cite{Dalgic:2006dt,Okamoto:2005zg,LCSR_04_BZ}. The dashed curves
on the phase shift plot show the effect on the statistical uncertainty
of reducing the input errors to $1/4$ of their current value. The
intercept of the phase shift with the horizontal line at $90^\circ$
indicates the position of a resonance.}
\label{fig:bpi}
\end{figure}

We use two subtraction points at $q^2=0$ and $\qsqmax=(m_B-m_\pi)^2$
and thus perform a three-parameter fit to $f_0(0)$, $f_0(\qsqmax)$ and
the scattering length $m_\pi a$. We find
\begin{equation}
f_0(0) = 0.257(31), \quad
f_0(\qsqmax) = 1.18(21), \quad
m_\pi a = 0.32(29).
\end{equation}
The fitted form factor and phase shift are shown in
Figure~\ref{fig:bpi}. We observe that the fitted value for
$f_0(\qsqmax)$ agrees within errors with the heavy quark effective
theory prediction in the soft-pion limit~\cite{Burdman:1993es},
$f_0(m_B^2) = f_B/f_\pi + \mathcal{O}(1/m_b^2) \approx 1.4(2)$ (using
$f_B=189(27)\mev$~\cite{Hashimoto:2004hn}). Our central phase-shift
curve shows evidence for a resonance at $\sqrt s \approx 5.6\gev$,
although we cannot give an upper bound for the resonance mass.

\subsection{Elastic $s$-wave $D\pi$ and $DK$ scattering}

To discuss the $D\pi$ phase shift we use
equation~(\ref{eq:hmchpt-amp}) with the obvious replacement $m_B\to
m_D$. For the $DK$ phase shift we project into the isospin zero
channel, where the two-particle irreducible amplitude again takes the
same form with the appropriate substitutions of masses and the
replacement $f_\pi \to f_K\approx 110\mev$.

We take input scalar form factor values from the Fermilab-MILC-HPQCD
lattice QCD calculation of reference~\cite{Aubin:2004ej}. The chiral
extrapolation procedure adopted there leads to parameters for a
Becirevic-Kaidalov (BK)~\cite{Becirevic:1999kt} parametrisation of
$f_0(q^2)$, and hence an explicit functional form, rather than values
at a set of $q^2$ points. We therefore generate a toy Monte Carlo
ensemble of BK parameters and minimise the integrated
squared-difference of the BK fit-function and a twice-subtracted
Omn\`es fit function to determine $f_0(0)$, $f_0(\qsqmax)$ and the
scattering length. We note that this fit could be avoided by using the
Omn\`es parametrisation throughout the analysis of the lattice data.

For the $D\pi$ case, we find a scattering length $m_\pi a = 0.29(4)$.
The output phase shift shows the existence of an $I=1/2$ $s$-wave
resonance at $2.2(1)\gev$.

For the $DK$ case, we find in almost all of our Monte Carlo trials
that the scattering length is huge, effectively infinite, telling us
that $\mathrm{Re}\,T^{-1}(\sth)=0$ as can be seen from
equation~(\ref{eq:Tinverse}). Hence there should be a resonance at
threshold, $(m_D+m_K)^2 = (2.36\gev)^2$. This can be understood by
noting the existence of a $0^+$ state, $D_{s0}^+(2317)$, discovered by
Babar~\cite{Aubert:2003fg}, which is likely an isoscalar~\cite{pdg}.
Neglecting isospin-violating decays to $D_s^+\pi^0$, this state could
be considered as an isoscalar $s$-wave $DK$ bound state. In this case,
following Levinson's theorem~\cite{MS70}, the phase shift close to
threshold has the form $\pi + p a +\cdots$, where $p$ is the
centre-of-mass three-momentum. Three-parameter fits (two subtractions
and $a$) show that the scattering length is effectively zero, so we
assume that the phase shift is $\pi$ over the range where the
integrand of the phase-shift integral is significant and obtain an
excellent two-parameter fit using two subtractions.

\section{$|V_{ub}|$ from exclusive semileptonic $B\to\pi$ decay}

\begin{figure}
\includegraphics[width=\hsize]{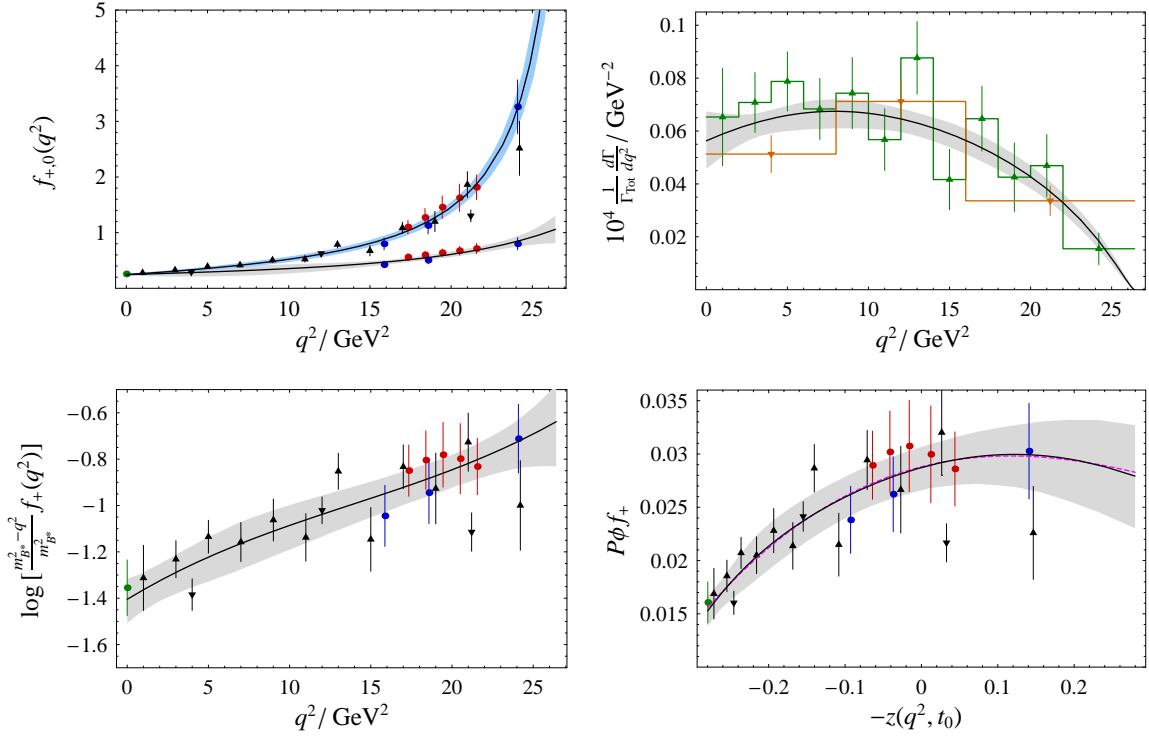}
\caption{Results obtained from the fit to experimental partial
  branching fraction data and theoretical form factor calculations.
  The top left plot shows the two form factors with their error bands,
  the lattice and LCSR input points (dots: green LCSR, red HPQCD, blue
  FNAL-MILC) and `experimental' points (black triangles,
  upward-pointing for tagged and downward pointing for untagged data)
  constructed by plotting at the centre of each bin the constant form
  factor that would reproduce the partial branching fraction in that
  bin. The top right plot shows the differential decay rate together
  with the experimental inputs. The bottom plots provide more details
  of the inputs and fits by showing on the left
  $\log[(\mbstar^2-q^2)f_+(q^2)/\mbstar^2]$ as a function of $q^2$,
  and on the right $P\phi f_+$ as a function of
  $-z$~\cite{Arnesen:2005ez,Becher:2005bg}. The dashed magenta curve
  in the bottom right plot is a cubic polynomial fit in $z$ to the
  Omn\`es curve.}
\label{fig:results}
\end{figure}

For our second application we use an Omn\`es representation for
$f(q^2)=(m_{B^*}^2-q^2)f_+(q^2)$ with many
subtractions~\cite{Flynn:2007ii,Flynn:2007qd} to motivate the
fit-function
\begin{equation}
f_+(q^2) = \frac1{\mbstar^2-q^2} \prod_{i=0}^n \left[
f_+(s_i) \big(\mbstar^2-s_i) \right]^{\alpha_i(q^2)}.
\end{equation}
We include $f_0$ information with a similar Omn\`es representation for
$f(q^2) =f_0(q^2)$ and apply the constraint $f_+(0)=f_0(0)$. This
provides an alternative to parametrisations based on the
$z$-expansion~\cite{Arnesen:2005ez,Becher:2005bg}. Adopting the fit
procedure described in~\cite{Flynn:2007qd}, we combine experimental
binned partial-branching fraction
information~\cite{Athar:2003yg,Hokuue:2006nr,Aubert:2006ry,Aubert:2006px}
for $f_+$ (to determine shape) with
lattice~\cite{Dalgic:2006dt,Okamoto:2005zg,Mackenzie:2005wu,VandeWater:2006aa}
and LCSR~\cite{LCSR_04_BZ} form-factor calculations of $f_+$ and $f_0$
(for normalisation and partial shape information). From a fit with
subtraction points at $\{0,1/3,2/3,1\}\qsqmax$, we determine:
\begin{equation}
\label{eq:best-fit}
\def\vubresult{(3.47\pm0.29) \times 10^{-3}}
\begin{array}{rcl}
\modvub         &=& \vubresult \\
f_+(0)=f_0(0)   &=& 0.245\pm0.023 \\
f_+(\qsqmax/3)  &=& 0.475\pm0.046 \\
f_+(2\qsqmax/3) &=& 1.07 \pm0.08
\end{array}\qquad
\begin{array}{rcl}
f_+(\qsqmax)    &=& 7.73 \pm1.29 \\
f_0(\qsqmax/3)  &=& 0.338\pm0.089 \\
f_0(2\qsqmax/3) &=& 0.520\pm0.041 \\ 
f_0(\qsqmax)    &=& 1.06 \pm0.26
\end{array}
\end{equation}
We also determine the combination $|V_{ub}|f_+(0) =
8.5(8)\times10^{-4}$ and the total branching fraction
\begin{equation}
\mathrm{B}(B^0\to\pi^- l^+\nu)=
 (1.37\pm0.08\pm0.01)\times10^{-4}
\end{equation}
where the first uncertainty is from our fit and the second is from the
uncertainty in the experimental $B^0$ lifetime. The result for
$\modvub$ is in striking agreement with $\modvub$ extracted using all
other inputs in CKM fits and shows some disagreement with $\modvub$
extracted from inclusive semileptonic $B\to\pi$ decays. In
Figure~\ref{fig:results}, we show our fitted form factor and
differential decay rate distribution.

\end{document}